\documentclass[fleqn,10pt]{wlscirep}
\usepackage[utf8]{inputenc}
\usepackage{amsmath,amssymb,amsfonts}%
\usepackage{amsthm}%
\usepackage{mathrsfs}%
\usepackage{graphicx}
\usepackage[T1]{fontenc}

\def \draft {1}

\usepackage{xparse}
\usepackage{ifthen}
\DeclareDocumentCommand{\comment}{m o o o o}
{\ifthenelse{\draft=1}{
    \textcolor{red}{\textbf{C : }#1}
    \IfValueT{#2}{\textcolor{blue}{\textbf{A1 : }#2}}
    \IfValueT{#3}{\textcolor{ForestGreen}{\textbf{A2 : }#3}}
    \IfValueT{#4}{\textcolor{red!50!blue}{\textbf{A3 : }#4}}
    \IfValueT{#5}{\textcolor{Aquamarine}{\textbf{A4 : }#5}}
 }{}
}

\newcommand{\todo}[1]{
\ifthenelse{\draft=1}{\textcolor{red!50!blue}{\textbf{TODO : \textit{#1}}}}{}
}

\title{ Ergodicity detection algorithms: Scaling of ergodicity in random symbolic dynamics  }

\author[1,*]{M. S\"uzen}

\affil[1]{APS}

\affil[*]{mehmet.suzen@physics.org}

\begin{abstract}
The mathematical definitions of distinct concepts that are needed in building an ergodicity detection
algorithm are introduced in a framework. This algorithmic framework is expressed in a discrete setting
in an accessible manner for broader quantitative practitioners without loss of generality. At the
same time, the common misconceptions of the requirement of visiting all available states in the
time-averaged quantities for physical systems and non-existence of an ergodic process are resolved
by introducing the distinction between Gibbs-Boltzmann and von Neumann-Birkhoff ergodic regimes. For
this purpose, we introduce a new concept which is called {\it sufficiency of sparse visit}.  We use
finite symbolic random sequences as a pedagogical tool in establishing the different approaches for
the detection of ergodic regimes of dynamical systems with vector patterns. The simple example system
conveys the different attitudes in ergodicity regimes and offers guidance for building computational
tools for its algorithmic detection.
\end{abstract}

\begin{document}

\flushbottom
\maketitle

\thispagestyle{empty}

\section{Introduction}\label{intro}

The origins of the {\it ergodic hypothesis} appear to be statements put forward by Boltzmann and Maxwell,
that multiple states evolving with constant energy should return to their initial conditions \cite{boltzmann64}.
This simple-sounding statement as the origin of Poincar{\'e} recurrence \cite{saussol09} formed the basis of
the connection between statistical mechanics and its thermodynamic justification, formalized by Gibbs ensemble
theory \cite{gibbs02}. However, the interpretation that a system should visit all possible states as a mapping
from constant energy  phases or from nearby points (quasi-ergodic) \cite{plato91} to its time evolution attracted the
interest of mathematical scientists \cite{lebowitz73}, i.e., von Neumann and Birkhoff via their
theorems \cite{birkhoff31, neumann32, moore15}, in some sense diverging from statistical physics to form a
mathematical field called {\it ergodic theory} for low dimensional systems as an extension of strong form of
law of large numbers \cite{choe05, lin07, mezard09}. In this context, the term {\it ergodicity}, a probe that measures if a
given dynamical regime attains ensemble averages being the same as time averages or if this equivalence is
broken, is studied as a key aspect of the dynamics of a physical system \cite{palmer82}.

Measuring ergodicity is practiced and needed in many different applied quantitative fields, such as, simple liquids \cite{mountain89},
the Ising-Lenz model with external fields and Hopfield networks \cite{suzen14, suzen16}, earthquake fault dynamics \cite{tiampo07},
atmospheric forecasting \cite{brener24}, economics \cite{peters19}, cognitive psychology \cite{hunter,molenaar}, financial
portfolios \cite{poitras15}, machine learning of optimal policies \cite{baumann25}, neuromorphic computing \cite{baccetti24},
random matrices \cite{suzen21sc, suzen20con}, thermal quantum dynamics \cite{kliczkowski24a, swietek24b, swietek25a}, 
deep learning architecture search \cite{suzen2017, suzen2019}, random binary sequences \cite{suzen2009}, 
artificial spin-ice experiments \cite{saccone23}, crystal phases \cite{sakak25}, relativistic geodesics \cite{suzen00}, 
quantum computing \cite{googleq25a}, Markov chains \cite{kingman61, pakes69, tweedie75} and  network congestion \cite{suzen08}. 
In biomedicine, sensitivity analysis in differential equations for disease trajectories would also require an attention 
to ergodicity \cite{suzen13}.

In the first section \ref{ergocondition}, we introduce the basic definition of ergodicity and distinctions we use for
different kinds of ergodicity via {\it sufficiency of sparse-visits}, the next section \ref{det} itemized the
details of conceptual components of detection algorithms in the discrete setting are defined, section \ref{sym} on experiments
presents numerical experiments with random symbolic dynamics demonstrating ergodicity detection framework in the mentioned two different
ergodicity schemes and their scaling properties. Finally, in the conclusion section \ref{conclusion}, we provide
final statements of the study.

\section{Gibbs-Boltzmann  ergodicity}
\label{ergocondition}

A mathematical ensemble, as introduced by Gibbs differentiate itself from probabilistic notion of
sample space. This was even noted in statistical inference literature \cite{mackay}, where ensemble
entails a more generic way of building set of possibilities. In the context of statistical mechanics,
this is the set of micro-states, i.e., mechanical degrees of freedoms. It might be obvious that
{\it von Neumann-Birkhoff  ergodicity} may not be attainable for thermodynamic systems due to
astronomical number of states to visit \cite{gaveau15}. However, {\it Gibbs-Boltzmann ergodicity}
states visiting most likely portion of the energy manifold in the microcanonical ensemble, this
brings the concept of {\it sufficiency of sparse visits}, that makes {\it ergodic
hypothesis} plausible. Moreover, von Neumann-Birkhoff  ergodicity would still hold for low-dimensional
systems \cite{deluca}.

We define a space that a system evolves over an ensemble $\Gamma_{i}^{j}$, with different
regions $i$ and formed with $j$ different states at that region. The regions are simply a collection of different states,
and not overlapping. This corresponds to all possible accessible microstates as established by Gibbs. A given
observable quantity $g$, as a function of ensemble members $g=g(\Gamma^{j})$, can be computed by taking its
{\it ensemble average}, $\langle g \rangle^{\gamma}$, at a given sparsity level $\gamma \in (0,1)$,
where 1.0 corresponds to no sparsity.  In the case of sparsity level 1.0, ensemble formed with all accessible states,
when some states merged and pooled regions can be formed then sparsity level is below 1.0, i.e., reduced representation
of the ensemble.  On the other hand, time-average of $g(t)$ is computed over states on the regions that system evolves
over time on the ensemble formed at the given sparsity level. The time-averaged $g$ at the given sparsity level is
expressed as,

$$ \langle g \rangle^{t}= \int_{t_{0}}^{t_{n}} d \Gamma_{i}^{\gamma} g(t).$$

Ergodicity of the system, or approach to ergodicity at given $\epsilon$ and sparsity, is measured by the
measure $\Delta$,

$$ \Delta_{\epsilon} =\Delta(\langle g \rangle^{\gamma},  \langle g \rangle^{t}) $$

$\Delta$ measure could be absolute difference for example, then approach to ergodicity at a given sparsity level
measured in terms of $\epsilon$,

$$ \epsilon^{\gamma} = | \langle g \rangle^{\gamma} -   \langle g \rangle^{t} | $$.

Gibbs-Boltzmann  ergodicity manifests at very large sparsities orders close to Avogadro's number
for macroscopic systems. This made the core of the concept of {\it sufficiency of sparse visits},
whereby very small number of visits are sufficient, compare to number of states that a system actually
can attain, this is due to energy manifolds values can be constructed with very small subsets of states
a system can attain \cite{boltzmann64, mountain89,gallavotti, dorfman99}. For very low dimensional systems, the
sparsity level of 1.0 corresponds to von Neumann-Birkhoff ergodicity \cite{deluca}. The concept
of {\it sufficiency of sparse visits} reflects the thermodynamics justification of ergodic hypothesis
Boltzmann and Gibbs established with contributions from Maxwell initially \cite{boltzmann64, gibbs02}.
Coarse-graining procedure is rather distinct in representing the degrees of freedom for the system before it
evolves over time, on the other hand {\it sufficiency of sparse visits} occurs over-time and it's the property
of dynamics that representation reduction occurs naturally. If a system reaches to {\it ergodic regime} in between
two time points $[t_{0}, t_{n}]$ at the acceptable $\epsilon$, then system said to be $ergodic$. In this sense,
a process is not ergodic rather a system attains the ergodic regime at a given time interval.

\begin{figure}[h!]
\centering
  \includegraphics[width=0.35\textwidth]{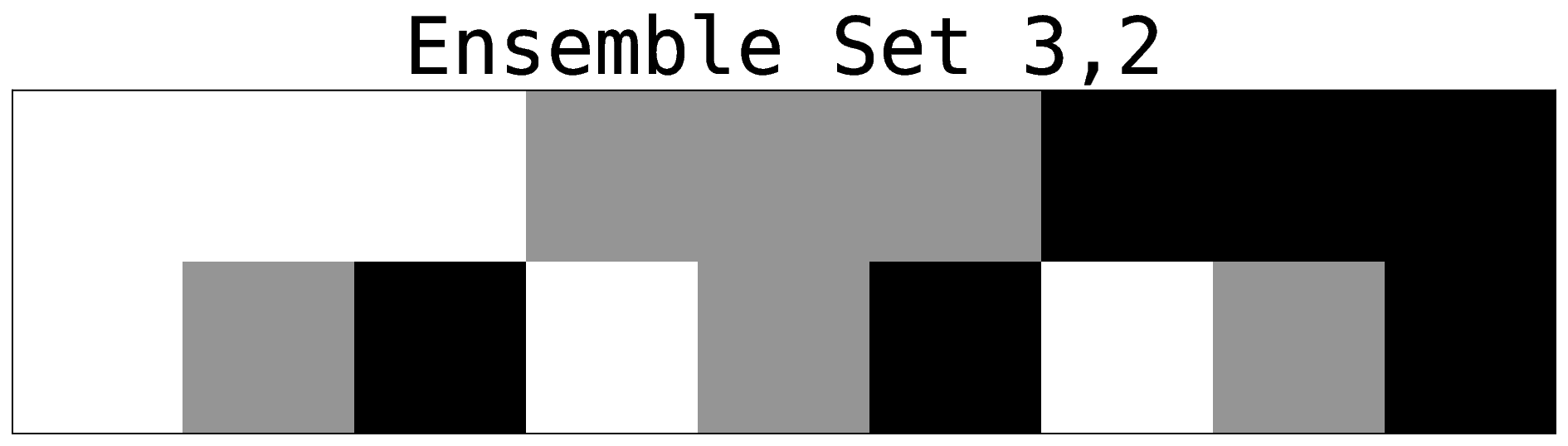}
  \caption{Discrete representation of an Ensemble set with alphabet size 3 and event size 2.
          Symbols are visualized with black, white and gray, columns are being ensemble members.}
  \label{ensembleset32}
\end{figure}

\begin{figure}[h!]
\centering
  \includegraphics[width=0.35\textwidth]{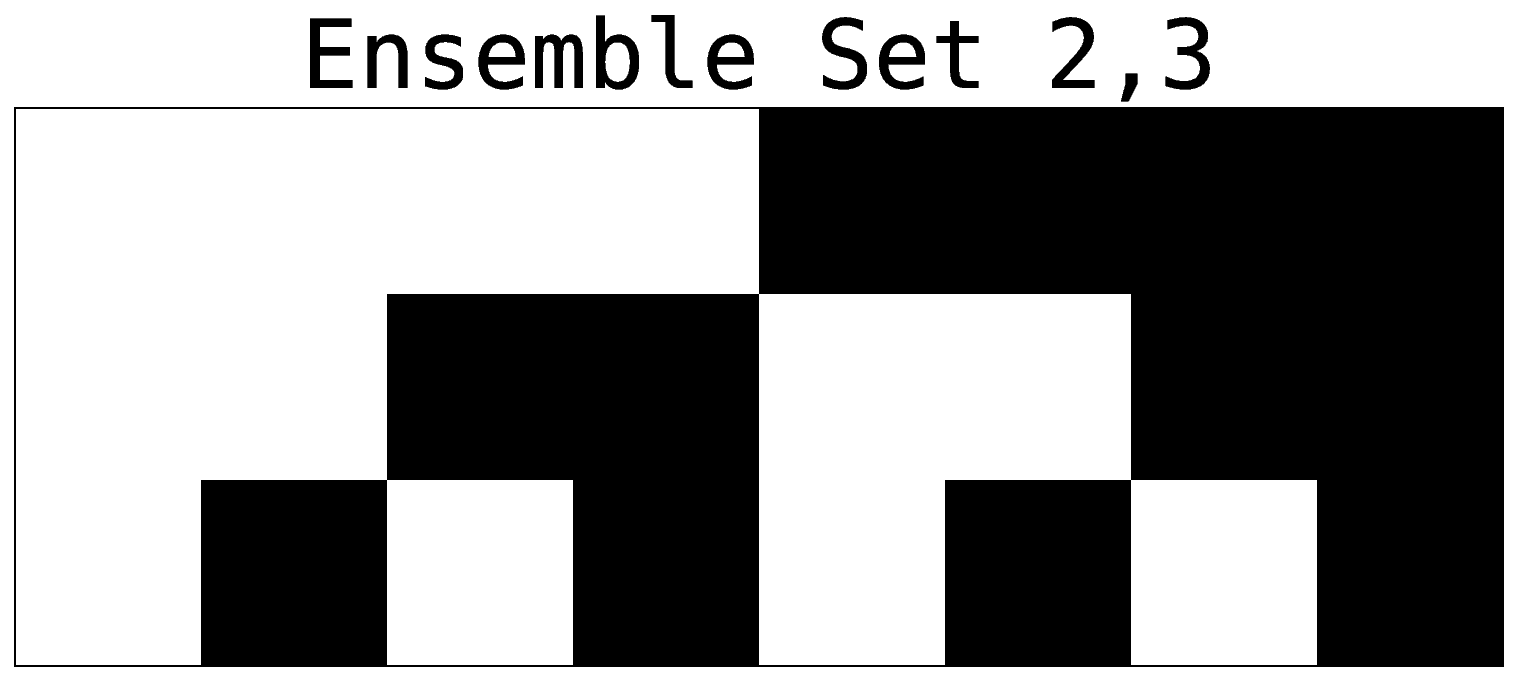}
  \caption{Discrete representation of an Ensemble set with alphabet size 2 and event size 3.
          Symbols are visualised with black and white, columns are being ensemble members.}
  \label{ensembleset23}
\end{figure}

\section{Ergodicity detection}
\label{det}

Discrete setting for a statistical mechanical system is not only pedagogically interesting \cite{hrabovsky20},
the study of dynamical system shown to be important in diverse scientific areas \cite{martelli11}. Because of this,
we build our framework of ergodicity detection in the language of discrete dynamical systems, using the
concepts introduced in the previous section.

We define a {\it discrete dynamical system} represented by the tuple for ergodicity,
$\langle \mathscr{S}, \mathscr{E}, \mathscr{U}, \mathscr{O}  \rangle$, where the set of states
$\mathscr{S} \in \mathbb{N}^{n}$, states can be restricted portion of $\mathscr{N}$, ensemble
$\mathscr{E}$ with members $e_{j}(\mathscr{S})$, set of dynamical rules $\mathscr{U}(t) = f_{i}(e_{j})$
with $f_{i}$ being functions of ensemble members appear as recurrence equations in evolving the system
from the initial states \cite{mackey} (dynamical law is a mapping from current states to the new points)
and an observable $\mathscr{O}$ as a function of ensemble members $e_{j}$.  If given dynamical rules
have a stochastic component then this will be random discrete dynamical system.

\begin{figure}[h!]
\centering
  \includegraphics[width=0.5\textwidth]{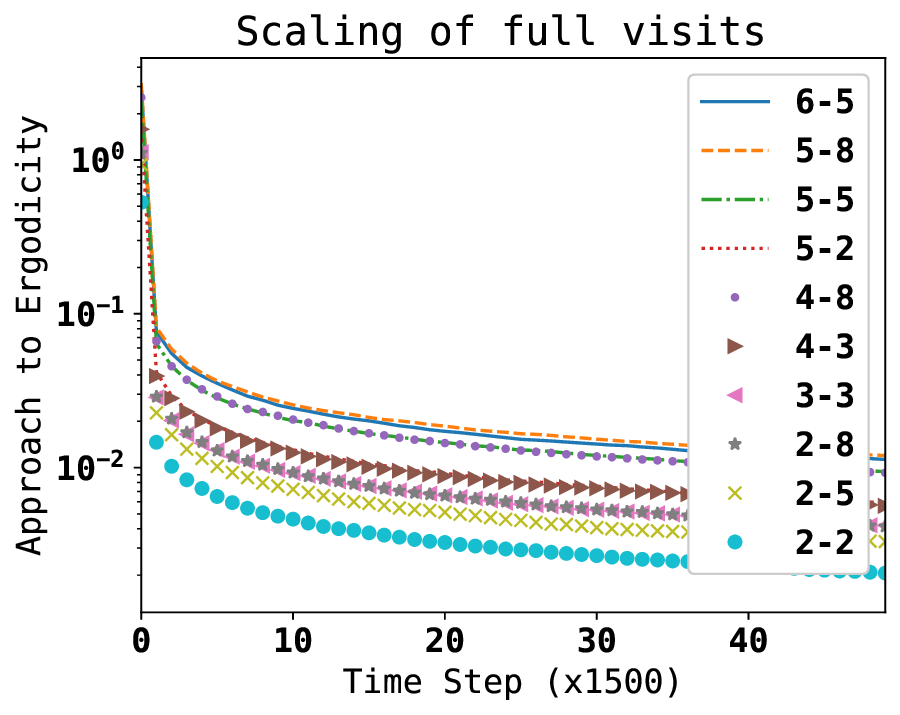}
  \caption{Approach to ergodicity with different alphabet and event sizes. A pattern that
           alphabet size and event size combined decimal number based approach to ergodicity
           scaling is observed.}
  \label{ergofull}
\end{figure}

{\small
\begin{table}[h!]
    \small
    \begin{tabular}{llllll}
      A & size & C & $C^{se}$ & $\alpha$ & $\alpha^{se}$ \\
      6 & 5 & 0.4997 & 0.0190 & -0.5059 & 0.0049 \\
      5 & 8 & 0.4834 & 0.0214 & -0.4952 & 0.0053 \\
      5 & 5 & 0.3633 & 0.0189 & -0.4921 & 0.0047 \\
      5 & 2 & 0.1765 & 0.0182 & -0.4936 & 0.0046 \\
      4 & 8 & 0.3831 & 0.0202 & -0.4946 & 0.0051 \\
      4 & 3 & 0.1677 & 0.0162 & -0.4944 & 0.0040 \\
      3 & 3 & 0.0601 & 0.0207 & -0.5024 & 0.0053 \\
      2 & 8 & 0.0429 & 0.0202 & -0.5001 & 0.0052 \\
      2 & 5 & -0.0848 & 0.0193 & -0.4913 & 0.0050 \\
      2 & 2 & -0.2310 & 0.0180 & -0.5051 & 0.0046 \\
   \end{tabular}
    \caption{Scaling of approach to ergodicity for different alphabet and event sizes, Fitting
            the power law over $\Omega$, approach to ergodicity, $C t^{\alpha}$, with standard
            error values identified with nested bootstrapping.}
    \label{tab:full}
\end{table}
}

\begin{figure}[h!]
\centering
  \includegraphics[width=0.5\textwidth]{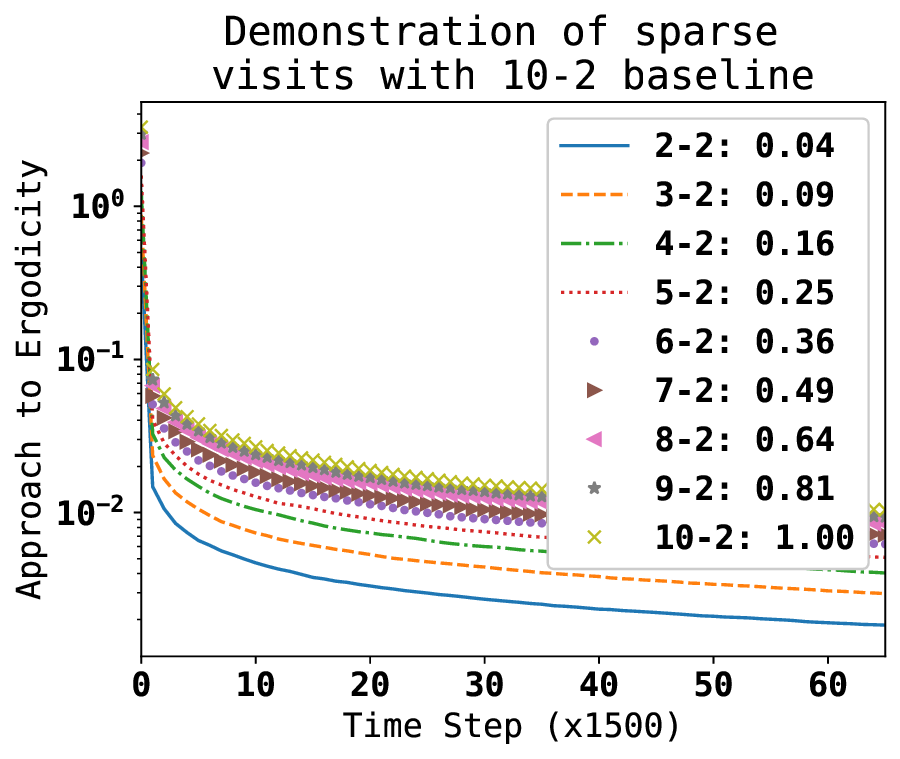}
  \caption{Using baseline of alphabet-event sizes 10-2, effect of deleting one symbol that corresponds to
           sparsity level, on the approach to ergodicity.}
  \label{ergosparse102}
\end{figure}

\begin{table}[h!]
    \small
    \begin{tabular}{lllllll}
      A & size & sparsity & C & $C^{se}$ & $\alpha$ & $\alpha^{se}$ \\
      2 & 2 & 0.0400 & -0.2555 & 0.0233 & -0.4974 & 0.0054 \\
      3 & 2 & 0.0900 & -0.0711 & 0.0219 & -0.4928 & 0.0051 \\
      4 & 2 & 0.1600 & 0.1109 & 0.0209 & -0.5021 & 0.0049 \\
      5 & 2 & 0.2500 & 0.1671 & 0.0208 & -0.4937 & 0.0049 \\
      6 & 2 & 0.3600 & 0.2791 & 0.0158 & -0.4988 & 0.0039 \\
      7 & 2 & 0.4900 & 0.3496 & 0.0213 & -0.5006 & 0.0048 \\
      8 & 2 & 0.6400 & 0.4222 & 0.0218 & -0.5008 & 0.0051 \\
      9 & 2 & 0.8100 & 0.4640 & 0.0219 & -0.5003 & 0.0052 \\
     10 & 2 & 1.0000 & 0.5164 & 0.0231 & -0.5014 & 0.0055 \\
   \end{tabular}
    \caption{Scaling of approach to ergodicity for different sparsities for the base case 10-2,
            the power law over $\Omega$, approach to ergodicity, $C t^{\alpha}$, with standard
            error values identified with nested bootstrapping.}
    \label{tab:102}
\end{table}

Ergodicity detection algorithm will have the following components,

\begin{enumerate}
  \item Set up the {\it discrete dynamical system} with sparsity level $\gamma$.
  \item Ensemble average of the observable denoted by $\langle \mathscr{O} \rangle$, computed over the
  ensemble $\mathscr{E}$.
  \item Let system evolve over time with an initial conditions, in the interval $[t_{0}, t_{n}]$ using the dynamical
  rules $\mathscr{U}(t)$, and obtain set of ensemble members $\mathscr{T}(t)$, this corresponds to sampling of ensemble
  with replacement obeying the dynamical rules. However,  $\mathscr{T}(t)$ forms a trajectory that it is an ordered
  set with temporal correlations,
  \item Time average of the observable denoted by $\langle \mathscr{O}(t) \rangle$ computed over the
  set $\mathscr{T}(t)$. This is repeated $n-$times, in order to compute mean time average $\langle \mathscr{O}(t) \rangle_{\mu}$.
  \item Set up an ergodicity measure $\Delta$ with $\epsilon$, deciding where system reaches to {\it ergodic regime}.
  \item Compute ergodicity over time $\Delta(t)$,  $m-times$ and find the mean,
  $$ \delta(t)^{\mu} = \Delta^{\mu}(\langle \mathscr{O} \rangle, \langle \mathscr{O}(t) \rangle_{\mu})$$
  \item If $\epsilon \ge \delta(t^{e})^{\mu}$, then the system reaches to an ergodic regime at time $t^{e}$. Obviously,
  if the condition is still satisfied with a monotonically decreasing $\delta^{\mu}$,after $t^{e}$ till $t^{f}$, then the system
  is in the ergodic regime in the
  interval $[t^{e}, t^{f}]$.
\end{enumerate}

If we track $\delta(t)^{\mu}$ over-time, we measure approach to ergodicity, denoted by $\Omega(t)$ for notational
convenience \cite{suzen14}. Note that by repeating trajectories $n \cdot m$-times, we essentially applied nested
bootstrap \cite{efron94}, which is necessary to get mathematically smooth estimate of the ergodicity measure over
time, i.e., approach to ergodicity curve.

\begin{figure}[h!]
\centering
  \includegraphics[width=0.5\textwidth]{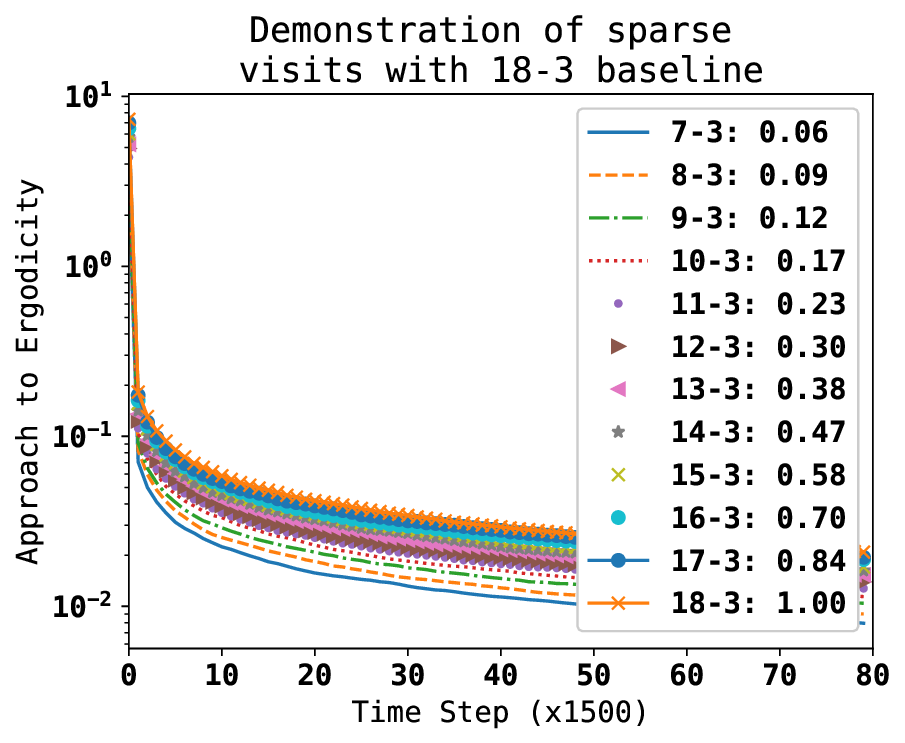}
  \caption{Using baseline of alphabet-event sizes 18-3, effect of deleting one symbol that corresponds to
           sparsity level, on the approach to ergodicity.}
  \label{ergosparse183}
\end{figure}

\begin{table}[h!]
    \small
    \begin{tabular}{lllllll}
      A & size & sparsity & C & $C^{se}$ & $\alpha$ & $\alpha^{se}$ \\
      7 & 3 & 0.0588 & 0.4259 & 0.0207 & -0.4971 & 0.0047 \\
      8 & 3 & 0.0878 & 0.5155 & 0.0211 & -0.5046 & 0.0048 \\
      9 & 3 & 0.1250 & 0.5298 & 0.0193 & -0.4955 & 0.0046 \\
     10 & 3 & 0.1715 & 0.6241 & 0.0227 & -0.5065 & 0.0053 \\
     11 & 3 & 0.2282 & 0.6273 & 0.0204 & -0.4977 & 0.0047 \\
     12 & 3 & 0.2963 & 0.6793 & 0.0205 & -0.5009 & 0.0047 \\
     13 & 3 & 0.3767 & 0.7014 & 0.0199 & -0.4965 & 0.0047 \\
     14 & 3 & 0.4705 & 0.7349 & 0.0203 & -0.4983 & 0.0047 \\
     15 & 3 & 0.5787 & 0.7838 & 0.0221 & -0.5023 & 0.0052 \\
     16 & 3 & 0.7023 & 0.7536 & 0.0208 & -0.4899 & 0.0048 \\
     17 & 3 & 0.8424 & 0.7956 & 0.0210 & -0.4935 & 0.0049 \\
     18 & 3 & 1.0000 & 0.8475 & 0.0215 & -0.4984 & 0.0049 \\
   \end{tabular}
    \caption{Scaling of approach to ergodicity for different sparsities for the base case 18-3,
            the power law over $\Omega$, approach to ergodicity, $C \Omega^{\alpha}$, with standard
            error values identified with nested bootstrapping.}
    \label{tab:183}
\end{table}

\section{Random symbolic dynamics}
\label{sym}

Working on a discrete setting for ergodicity detection naturally bring the discrete dynamical
systems and consideration of symbolic dynamics \cite{morse38}. In essence, we operate on and analyze
the sequences generate by an alphabet $\mathscr{A}$, which forms the states $\mathscr{S}$ of our
dynamical system. We can build ensembles over symbols using different shift maps \cite{lindmarcus}.
The simplest example in this context is probably a Bernoulli process, states $\mathscr{S}=\{0, 1\}$.
An ensemble of sequence of length 3 for these states can be represented as black and white cells. Then the ensemble
looks like in Figure \ref{ensembleset23}, columns being ensemble members, the set
$\{000, 100, 010, 110, 001, 101, 011, 111\}$. This can be generated via Cartesian products on the
alphabet repeatedly by the sequence length. Using NumPy \cite{numpy}, we utilize state-of-the-art
random generator called PCG \cite{pcg}, implementing ergodicity detection algorithm for
random symbolic dynamics \cite{suzen25ergo}.

Similarly, we show the ensemble of  $\mathscr{S}=\{0, 1, 2\}$ with sequence of length of 2,
color mapped white, black and gray in Figure \ref{ensembleset32}.

The following components are set for random symbolic dynamics:

\begin{enumerate}
\item The tuple $\langle \mathscr{S}, \mathscr{E}, \mathscr{U}, \mathscr{O}  \rangle$ set as follows.
      The alphabet or states $\mathscr{S}$ are formed with the set of positive integers and zero. The ensemble
      set is formed via repeated Cartesian product given sequence size
      $\mathscr{E}= \mathscr{S}\times \mathscr{S} ... \times \mathscr{S}$. The function to generate
      trajectories is simply a random choice $\mathscr{R}$ with replacement over $\mathscr{E}$ repeatedly,
      $\mathscr{U} = \mathscr{R}(\mathscr{E})$. This is simplified dynamics without explicit recurrence.
      Observable is simply the sum of the given ensemble element $e_{j}$.
\item Ensemble average is computed by taking mean of the all sums of the ensemble elements.
      We apply given sparsity to reduce the size of ensemble if $\gamma < 1.0$.
\item We sample the ensemble 120000 times with $\mathscr{R}$, to get a single trajectory.
      40 trajectories are generated. Mean observable over trajectory is identified, i.e., sum of
      each ensemble member sampled.
\item Ergodicity measure, absolute difference between time and ensemble average is used.
      This is repeated 70-times.
\item We found the power law \cite{newman}, $\Omega = C t^{\alpha}$, for approach to ergodicity. $\alpha$ is the
      exponent for the scaling law and $C$ measures the strength of the approach to ergodicity.
      Both $\alpha$ and $C$ are computed 70 times, reporting their mean and standard errors.
\end{enumerate}

For generic ergodicity detection algorithms, nested bootstrap repeats may be different.

We compute approach to ergodicity with no sparsity, corresponding to the von Neumann-Birkhoff
ergodicity, for different alphabet sizes and system lengths: 65, 58, 55, 52, 48, 43, 33, 28, 25, 22.
Shown in Figure \ref{ergofull}. The rate of convergence has a smooth ordering based on alphabet size
and sequence lengths. Measuring strength of the power law has similar ordering as reported in
Table \ref{tab:full}. Doubling the alphabet length approximately doubles the strength but we
see a constant scaling exponent over all lengths, about 0.5. This constant power-law scaling indicates
random nature.

In order to simulate and find the effect of sparsity, i,e., Gibbs-Boltzmann ergodicity
for the principle of {\it sufficiency of sparse visits}: We set a baseline systems where by
we consider its full ensemble with sparsity level $\gamma=1.0$. Its sparse versions are
determined by the ensemble size of smaller systems. We computed ergodicity for
two baseline systems $10-2$ and $18-3$, their alphabet and sequence sizes. Results are
reported on Figures \ref{ergosparse102} - \ref{ergosparse183} and Tables \ref{tab:102}-\ref{tab:183}.
We observe similar smooth variation of power-law strengths over smaller sparsity. It is shown that
order of magnitude sparsity made approach to attainment of ergodicity much faster. This indicates,
very large sparsities should occur in order to attain ergodic regimes for macroscopic systems.

\section{Conclusions}
\label{conclusion}

In this work we provide a clear distinction between {\it von Neumann-Birkhoff} and {\it Gibbs-Boltzmann}
ergodicities reflecting the conditions where {\it ergodic hypothesis} can be attained. Along these lines,
measuring and detecting ergodic regime algorithmically is formalized for the development of family
of {\it ergodicity detection algorithms}. This formalization and understanding necessary components
would help practitioners in diverse fields. We have demonstrated application of the frameworks on the
random symbolic dynamics, using diverse systems and identify scaling of ergodicity in these systems.

\section*{Acknowledgments}
Y. S\"uzen for her kind support and encouragements.

\section*{Declarations}
\begin{itemize}
\item The author declare that he has no conflict of interest.
\item There is no data release with the work. 
\item Code is released under a public repository \cite{suzen25ergo}.
\end{itemize}

\bibliography{suzen}

\end{document}